\documentclass[prd,showpacs]{revtex4}
\usepackage{graphicx}
\begin{document}
\title{
Magnetic field effects on neutrino oscillations}
\author{Andrea Erdas}
\email{aerdas@loyola.edu}
\author{Zachary Metzler}
\email{zbmetzler@loyola.edu}
\affiliation{
 Department of Physics, Loyola University Maryland 4501 N Charles Street,
       Baltimore, MD 21210}
\date{July 4, 2019}
\begin {abstract} 
Using the exact propagators
in a constant magnetic field, the neutrino self-energy has been calculated
to all orders in the field strength $B$ within the minimal extension of the Weinberg-Salam model
with massive Dirac neutrinos. A neutrino dispersion relation, effective potential and effective mass have been obtained that depend on 
$B$. The consequences of this effective potential on neutrino oscillations have been explored, and resonance conditions have been obtained for magnetic fields $0<eB
\le 4m^2_e$, where $e$ is the elementary charge and $m_e$ is the electron mass. The oscillation length has also been obtained, showing that $\nu_e$-$\nu_\mu$ resonant oscillations 
in magnetic field are likely to occur in the proximity of most stars generating high magnetic fields.
\end {abstract}
\maketitle
\section{Introduction}

The study of neutrinos in the presence of a magnetic field is important in astrophysics and cosmology~\cite{raffeltbook,giuntibook}. The neutrino
self-energy and its dispersion relation are modified in a magnetic field, and such modifications have been studied extensively in the 
literature~\cite{erdaskim,dolivo,elmfors,erdasfeld,erdas,kuznetsov2,mckeon,elizalde1,elizalde2,Nieves:2018qwg}.
There is a  natural magnetic field scale that is required to significantly impact
quantum processes, dependent on the electron mass $m_e$ and the elementary charge $e$, 
$B_e=m^2_e/e = 4.41 \times 10^{13}$ G. Large magnetic fields are present in many 
astrophysical sites such as supernovae, neutron stars, and white dwarfs. Fields larger than
$B_e$ can appear during the explosion of a supernova  or the coalescing of a neutron star. 
Magnetars are young neutron stars that generate even stronger magnetic fields, $10^{14}-10^{16}$ G \cite{Duncan,Thompson1,Thompson2}.
Magnetic fields as high as
$10^{22}-10^{24}$ G  have been hypothesized to exist during the electroweak
phase transition of the early universe \cite{Brandenburg,Joyce,grasso}. Although it is rare for neutrinos to encounter magnetic fields larger than $10^{16}$ G, in many 
situations neutrinos come upon astrophysical sites where magnetic field
strengths are at or around $B_e$. 

While the electromagnetic properties of massless neutrinos have been studied extensively and for a long time~\cite{erdaskim,dolivo,elmfors,erdasfeld}, the nontrivial electromagnetic properties of massive neutrinos have only been addressed more recently \cite{Giunti:2014ixa,Studenikin:2018vnp,Tarazona:2017jnd}. A recent investigation on the role of a neutrino magnetic moment in flavor, spin and spin-flavor oscillations has produced interesting results \cite{Popov:2018seq,Popov:2019nkr}, but does not consider magnetic effects beyond those linear in $B$.
In  this paper we will focus on massive Dirac neutrinos within the minimally extended Standard Model and take a rigorous field theoretical approach to calculating their self-energy in a homogeneous magnetic field. We will accomplish this task by using Schwinger's proper time method \cite{schwinger} and the 
exact fermion, scalar, and $W$-propagator in a constant magnetic field \cite{erdasfeld,erdas}.

In Section \ref{2} the notation for the fermion, gauge boson and scalar propagators in magnetic field is reviewed
and the one-loop neutrino self-energy is set up. In Section \ref{3} we calculate the self-energy operator and obtain a simple expression for the self-energy in the limit of $eB\ll m^2_W$, where $m_W$ is the mass of the $W$-boson, exact up to order $B^2$. In Section \ref{4} we use the self-energy operator we obtained to calculate the dispersion relation, effective potential and effective mass for  massive Dirac neutrinos in a magnetic field and show that terms in the self-energy of order $B^2$ are often dominant when compared to linear terms such as the magnetic dipole moment term. In Section \ref{5} we use the effective potential to investigate neutrino oscillation in a magnetic field and obtain resonance condition and oscillation length.
\section{ Propagators and neutrino self-energy in a constant magnetic field}
\label{2}
The metric used in this paper is
$g^{\mu \nu} = \mathrm{diag}(-1,+1,+1,+1)$ and the $z$-axis points in the direction of the constant 
magnetic field $\mathbf{B}$. Therefore the electromagnetic
field strength tensor $F^{\mu \nu}$ has only two non-vanishing components $F^{12}= -F^{21} = B$. 

For the purpose of this work, it would seem convenient to work in the
unitary gauge where the unphysical scalars
disappear. 
However, the $W$-propagator is quite cumbersome in this gauge, and we prefer
to work in the Feynman gauge, where the $W$-propagator has a much simpler expression. The
following expressions for the charged lepton
$S(x',x'')$ \cite{schwinger}, $W$-boson
$G^{\mu \nu}(x',x'')$ and scalar propagators $D(x',x'')$  \cite{erdasfeld} in a constant
magnetic field have been written using Schwinger's proper time method:
\begin{equation}
S(x',x'')=\Omega^*(x',x'')\!\int{d^4k\over (2\pi)^4}e^{ik\cdot (x'-x'')}
S(k) \quad , 
\label{sx}
\end{equation}
\begin{equation}
G^{\mu\nu}(x',x'')=\Omega(x',x'')\!\int{d^4k\over (2\pi)^4}e^{ik\cdot
(x'-x'')}
G^{\mu\nu}(k) \quad ,
\label{gx}
\end{equation}
\begin{equation}
D(x',x'')=\Omega(x',x'')\!\int{d^4k\over (2\pi)^4}e^{ik\cdot (x'-x'')}
D(k) \quad .
\label{dx}
\end{equation}
In the Feynman gauge, the translationally 
invariant parts of the propagators are
\begin{equation}
\label{S}
S(k) =\int_0^\infty \!\!{ds\over\cos eBs} 
{\exp{\left[-is\left(m^2_\ell+k^2_{\parallel}+
k^2_{\perp}{\tan eBs\over eBs}\right)\right]}}
\left[(m_\ell-
\not\! k_{\parallel})e^{-ieBs\sigma_3}-
{\not\! k_{\perp}\over \cos eBs}
\right] , 
\end{equation}
\begin{equation}
G^{\mu \nu}(k)=
\int_0^\infty \!\!{ds\over\cos eBs}\,
{\exp{\left[-is\left(m^2_W+k^2_{\parallel}+
k^2_{\perp}{\tan eBs\over eBs}\right)\right]}}
[g^{\mu \nu}_{\parallel}
+(e^{2eFs})^{\mu\nu}],
\label{G}
\end{equation}
\begin{equation}
D(k)=
\int_0^\infty \!\!{ds\over\cos eBs}\,
{\exp{\left[-is\left(m^2_W+k^2_{\parallel}+
k^2_{\perp}{\tan eBs\over eBs}\right)\right]}},
\label{D}
\end{equation}
where $-e$ and
$m_\ell$ are the charge and mass of the charged lepton $\ell$, and
$m_W$ is the $W$-mass. It is convenient to use the notation
\begin{equation}
a^\mu_{\parallel}=(a^0,0,0,a^3), \quad
a^\mu_{\perp}=(0,a^1,a^2,0)
\label{amu}
\end{equation}
and 
\begin{equation}(a b)_\parallel=-a^0\,b^0+a^3\,b^3, \quad
(a b)_\perp=a^1\,b^1+a^2\,b^2
\label{adotb}
\end{equation}
for arbitrary four-vectors $a$ and $b$. Using this notation we write the metric tensor as
\begin{equation}
g^{\mu\nu}=g^{\mu\nu}_\parallel+g^{\mu\nu}_\perp .
\label{g}
\end{equation} 
The $4\times 4$ matrix $\sigma_3$ that appears in the charged lepton propagator (\ref{S}), 
is
\begin{equation} 
\sigma_3=
{i \over 2}[\gamma^1, \gamma^2].
\label{sigma3}
\end{equation} 
When writing the $W$-propagator (\ref{G}), we use the notation
\begin{equation}
\left(e^{2eFs}\right)^{{\mu}\nu}=
g^{\mu\nu}_{\perp} \cos{(2eBs)}
+{F^{\mu\nu}\over B} \sin{(2eBs)} \quad .
\label{efmunu}\end{equation}
We choose the electromagnetic vector potential to be
$A_\mu=-{1\over2}F_{\mu \nu}x^\nu$ and therefore
the phase factor which appears in Eqs.  (\ref{gx}), (\ref{dx}) is given by
\begin{equation}
\Omega(x',x'')=\exp\left(
i {e\over 2}x''_\mu F^{\mu \nu} x'_\nu
\right)  \quad ,
\label{2_08a}
\end{equation}
with $\Omega^*(x',x'')$ in Eq. (\ref{sx}) being its complex conjugate.

Perturbatively, the self-energy operator in the Feynman gauge corresponds to the sum of two 
diagrams, a bubble diagram with the gauge boson and a bubble diagram with the scalar
\begin{equation}
\Sigma(p)=\Sigma_W(p)+\Sigma_\Phi(p).
\label{Sigma}
\end{equation}
The translationally non-invariant phase factors $\Omega(x',x'')$ are identical for all propagators
and the product of phase factors in the two-vertex loop is
\begin{equation}
\Omega^*(x',x'')\Omega(x',x'')=1,
\label{phiprod}
\end{equation}
therefore, within the minimally extended version of the standard model of
electroweak interactions with an $SU(2)$-singlet right-handed neutrino, the two bubble diagrams can be written as \cite{erdasfeld,kuznetsov3}
\begin{equation}
\Sigma_W(p)=-i{g^2\over 2}R{\gamma}_{\alpha}
\int{d^4k\over (2\pi)^4}S(p-k)G^{\beta\alpha}(k){\gamma}_{\beta}L,
\label{sigmaw}
\end{equation}
\begin{equation}
\Sigma_\Phi(p)=-i{g^2\over 2m^2_W}[m_\ell R-m_\nu L]
\int{d^4k\over (2\pi)^4}S(p-k)D(k)[m_\ell L-m_\nu R],
\label{sigmaphi}
\end{equation}
where $g$ is the $SU(2)$ coupling constant, $L={1\over 2}(1-\gamma_5)$ and
$R={1\over 2}(1+\gamma_5)$ are the left-handed and right-handed projectors
and neutrino mixing is allowed by taking a nondiagonal neutrino mass matrix
$m_\nu$ in Eq. (\ref{sigmaphi}).
\section{The self-energy operator}
\label{3}

Inserting the expression for the propagators from Eqs. (\ref{S}) and (\ref{G}) 
into the self-energy, we write ${\Sigma}_{W}(p)$ as
\begin{eqnarray}
\Sigma_{W}(p)=&&-{ig^2\over 2}
\int\!{{d^4 k}\over{{(2{\pi})}^4}}{\int}^{\infty}_{0}
{ds_1\over{\cos z_1}}{\int}^{\infty}_{0}
{ds_2\over{\cos z_2}}{e^{-is_1(m^2_\ell +q^2_{\parallel}
+q^2_{\perp}{{\tan {z_1}}\over {z_1}})}}
{e^{-is_2(m_W^2 +k^2_{\parallel}+k^2_{\perp}{{\tan {z_2}}\over
{z_2}})}}{\times}
\nonumber \\
&&R{\gamma}_{\alpha}
\left[(m_\ell-
\not\! q_{\parallel})e^{-iz_1\sigma_3}-
{\not\! q_{\perp}\over \cos z_1}
\right]
[g^{\beta\alpha}_{\parallel}
+(e^{2eFs_2})^{\beta\alpha}_{\perp}]{\gamma}_{\beta}L
\end{eqnarray}
where
\begin{equation}
q = p-k \quad , \quad\quad z_1 = eBs_1 \quad , \quad\quad z_2 = eBs_2 \quad
.
\end{equation}
We do the straightforward $\gamma$-algebra, change variables from $s_i$
to
$z_i$, translate the $k$ variables
of integration 
and do the four gaussian integrals over the shifted variables
$k$, to obtain
\begin{eqnarray}
\Sigma_{W}(p)=&& {g^2\over (4\pi)^2}
{\int}^{\infty}_{0}
{\int}^{\infty}_{0}
{dz_1 dz_2 \over (z_1+z_2) \sin(z_1+z_2)}
e^{-i[z_1m^2_\ell + z_2 m_W^2  + {\cal P}]/eB}  \times
\nonumber \\
&& \left[{z_2\over z_1+z_2}{\not\! p_{\parallel}}e^{
i\sigma_3(z_1+2 z_2)}+{\sin z_2 \over \sin(z_1+z_2)}
{\not\! p_{\perp}}\right]
\!L + (\textrm{c.t.})_W
\label{sigmaw2}
\end{eqnarray}
where
\begin{equation}
{\cal P}={z_1 z_2 \over (z_1 + z_2)}
p_{\parallel}^2
+{\sin{z_1} \sin{z_2} \over \sin{(z_1+z_2)}}
p_{\perp}^2 \quad ,
\end{equation}
and the appropriate counter-terms (c.t.) are added.
Next it is convenient to change integration
variables from $(z_1, z_2)$ to $(s, u)$ defined by
\begin{equation}
z_1 = eBs u=zu\quad\quad \textrm{and} \quad\quad
z_2=eBs (1-u) =z(1-u)\quad .
\end{equation}
The result is
\begin{equation}
\Sigma_{W}(p)={g^2\over (4\pi)^2} {\int}^{\infty}_{0}{dz\over\sin z} {\int}^{1}_{0}du\,\,e^{-is\Lambda^2}
R\left[(1-u)e^{iz(2-u)\sigma_3}{\not\! p_{\parallel}}+{\sin [z(1-u)] \over \sin z}{\not\! p_{\perp}}\right]\!L 
+ (\textrm{c.t.})_W
\label{sigmaw3}
\end{equation}
where
\begin{equation}
\Lambda^2= um^2_l+(1-u)m^2_W+u(1-u)p_{\parallel}^2+{\sin zu \sin (z-zu)\over z \sin z}p_{\perp}^2.
\end{equation}
The scalar-loop contribution to the self-energy is found similarly
\begin{equation}
\Sigma_{\Phi}(p)={g^2\over (4\pi)^2} {\delta_l\over 2}{\int}^{\infty}_{0}{dz\over\sin z} {\int}^{1}_{0}du\,\,e^{-is\Lambda^2}
R\left[(1-u)e^{-izu\sigma_3}{\not\! p_{\parallel}}+{\sin [z(1-u)] \over \sin z}{\not\! p_{\perp}}\right]\!L 
+ (\textrm{c.t.})_\Phi
\label{sigmaphi3}
\end{equation}
where we used
\begin{equation}
\delta_l= {m^2_l\over m^2_W},
\end{equation}
and neglected the matrix $m_\nu$ from  Eq. (\ref{sigmaphi}) whose elements are all of the order of $2{\rm eV}$ or less.

At this point we focus on magnetic fields $eB\ll m^2_W$, since magnetic fields $eB\sim m^2_W$ are not found in the universe.
The expression of  $\Sigma_{W}(p)$ in the limit of $eB\ll m^2_W$ is readily found by taking $z\ll1$, and we obtain
\begin{eqnarray}
\Sigma_{W}(p)=&&{g^2\over (4\pi)^2} {\int}^{\infty}_{0}{ds\over s} {\int}^{1}_{0}du\,\,e^{-is\Lambda^2_0}(1-u)
R\biggl[i(2-u){z\over m^2_W}\sigma_3{\not\! p_{\parallel}}+{z^2\over 6 m^4_W}{\not\! p}-{z^2\over 2 m^4_W}(2-u)^2{\not\! p}_{\parallel}
\nonumber \\
&&   +{z^2\over 6 m^4_W}u(2-u){\not\! p}_{\perp}-is{z^2\over 3 m^4_W}u^2(1-u)^2\left({p^2_\perp\over m_W^2}\right){\not\! p}\biggr]\!L ,
\label{sigmaw4}
\end{eqnarray}
where we neglected terms that do not depend on the magnetic field since they will be absorbed by the neutrino wavefunction and mass renormalization, and we used the notation
\begin{equation}
\Lambda^2_0= (1-u)+u{m^2_l\over m^2_W}+(1-u)u{p^2\over m^2_W}.
\end{equation}
Since $p^2\sim m^2_\nu$, where $m_\nu$ is the mass of the neutrino, we can take $\Lambda^2_0\simeq (1-u)+u\delta_l$ and, after doing the $s$ and $u$ integration, we obtain
\begin{eqnarray}
\Sigma_{W}(p)=&&{g^2\over (4\pi)^2} 
R\Biggl[
{3eB\over 2m^2_W}\sigma_3{\not\! p_{\parallel}}
+{(eB)^2\over 6 m^4_W}\left({5\over 2} +{4p^2_\perp\over 3m_W^2}+\ln \delta_l\right){\not\! p}
\nonumber \\
&& +{(eB)^2\over 2 m^4_W}\left(1 - \ln \delta_l\right){\not\! p}_{\parallel}   
+{(eB)^2\over 6 m^4_W}\left(\ln \delta_l\right){\not\! p}_{\perp}\Biggr]\!L.
\end{eqnarray}
Undertaking similar steps, we find
\begin{eqnarray}
\Sigma_{\Phi}(p)=&&{g^2\over (4\pi)^2} \delta_l
R\Biggl[
-{eB\over 4m^2_W}\sigma_3{\not\! p_{\parallel}}
+{(eB)^2\over 3 m^4_W}\left({1\over 4} +{p^2_\perp\over 3m_W^2}+{1\over 4}\ln \delta_l\right){\not\! p}
\nonumber \\
&& -{(eB)^2\over 4 m^4_W}\left({5\over 2} + \ln \delta_l\right){\not\! p}_{\parallel}   
+{(eB)^2\over 12 m^4_W}\left({3\over 2}+\ln \delta_l\right){\not\! p}_{\perp}\Biggr]\!L,
\end{eqnarray}
where each term is suppressed by a factor of $\delta_l$ when compared to the similar term in $\Sigma_W(p)$, thus making $\Sigma_\Phi(p)$ negligible when compared to $\Sigma_W(p)$. The self-energy operator is therefore $\Sigma(p)\simeq \Sigma_W(p)$.

\section{Dispersion relation, effective potential and effective mass}
\label{4}
We intend to interpret the effect of the magnetic field on the self-energy operator as an effective neutrino mass \cite{Bethe:1986ej,erdasfeld}. To do so we must first evaluate the average value of $\Sigma(p)$ over the neutrino spinor in the mass basis,  $ \langle \Sigma \rangle = \bar{u}_i\Sigma(p) u_i$ , since neutrino mass is defined in the mass basis and not in the flavor basis.
We find,
\begin{equation}
\bar{u}_i R \sigma_3{\not\! p_{\parallel}} L u_i=- {m_i s_3\over 2}
\label{u1}\end{equation}
where $s_3=\pm1$, depending on the neutrino spin orientation relative to the magnetic field and, similarly,
\begin{equation}
\bar{u}_i R{\not\! p} L u_i=- {m_i\over 2},
\label{u2}\end{equation}
and
\begin{equation}
\bar{u}_i R{\not\! p}_{\perp} L u_i= \chi\bar{u}_i {\not\! p}_{\perp} u_i,
\label{u3}\end{equation}
where $\chi={1\over 2}\langle 1-{\vec{\sigma}\cdot\mathbf{p}\over E+m_i} \rangle $ in the Dirac representation and $E$ and $\mathbf{p}$ are the neutrino energy and momentum.
Indicating with $U$ the neutrino mixing matrix
\begin{equation}
\nu_\ell=\sum_{i=1}^3U_{\ell i}\nu_i,
\label{Umix}\end{equation}
and using Eqs. (\ref{u1}), (\ref{u2}), and (\ref{u3}), we write
\begin{equation}
\langle\Sigma\rangle=-{g^2\over 64\pi^2} {eB\over m^2_W}\sum_{\ell}U_{\ell i}^\star U_{\ell i}
\Biggl[
3s_3m_i
+{eB\over 3 m^2_W}\left({11\over 2} +{4p^2_\perp\over 3m_W^2}-2\ln \delta_\ell\right)m_i
 +{2eB\over  m^2_W}\left(1 -{4\over 3}\ln \delta_\ell\right)\chi\bar{u}_i {\not\! p}_{\perp} u_i\Biggr],
\label{sig}\end{equation}
where $\ell = e,\mu,\tau$. Eq (\ref{sig}) agrees with \cite{erdasfeld} in the absence of mixing and neutrino masses. Notice that transitions of the type  $\nu_i\rightarrow \nu_j$ are suppressed \cite{Borisov:1991cp} by a GIM type mechanism, and therefore quantities such as $\bar{u}_j\Sigma(p) u_i$ can be neglected. We obtain the dispersion relation by setting 
\begin{equation}
{\not\! p} +m_i - \langle\Sigma\rangle=0,
\end{equation}
and find the following energy-momentum relation for the massive neutrino in a magnetic field
\begin{equation}
E^2=\mathbf{p}^2+2\Lambda p^2_\perp+(1+2\mu)m_i^2,
\label{dispersion}
\end{equation}
with
\begin{equation}
\Lambda ={g^2\over 32\pi^2} {(eB)^2\over m^4_W}\sum_{\ell}U_{\ell i}^\star U_{\ell i}\left(1 -{4\over 3}\ln \delta_\ell\right)\chi,
\label{Lambda}\end{equation}
and
\begin{equation}
\mu ={g^2\over 64\pi^2} {eB\over m^2_W}\sum_{\ell}U_{\ell i}^\star U_{\ell i}
\Biggl[
3s_3
+{eB\over 3 m^2_W}\left({11\over 2} +{4p^2_\perp\over 3m_W^2}-2\ln \delta_\ell\right)
\Biggr].
\label{mu}\end{equation}
Eq. (\ref{mu}) is valid in the case of weak ($eB\ll m^2_e$) and moderate ($m^2_e\ll eB\ll m^2_W$) magnetic field, while Eq. (\ref{Lambda}) is valid only in the case of weak magnetic
field. In the case of moderate magnetic field $\Lambda$ becomes \cite{erdas}
\begin{equation}
\Lambda ={g^2\over 32\pi^2} {(eB)^2\over m^4_W}\sum_{\ell}U_{\ell i}^\star U_{\ell i}\left[1 -{4\over 3}\ln\left({eB\over 6m^2_W} +\delta_\ell\right)\right]\chi,
\label{lambda2}\end{equation}
which is more general than Eq. (\ref{Lambda}) and renders obsolete the distinction between weak and moderate magnetic field.

The dispersion relation of Eq (\ref{dispersion}) yields the following effective neutrino mass
\begin{equation}
m_{eff}=\sqrt{(1+2\mu)m_i^2+2\Lambda p^2_\perp}.
\label{mu2}\end{equation}
Notice that, since $m_i\le 2 {\rm eV}$, for $p_\perp \sim 1 {\rm MeV}$ or higher the $\Lambda$ term in Eq. (\ref{mu2}) dominates over the $\mu$ term, which can be safely neglected in many situations.

The effective potential is most useful in the flavor basis and, neglecting terms similar to $\mu$ of Eq. (\ref{mu}) for the reason stated above, we find
\begin{equation}
V_\ell=\lambda_\ell p^2_\perp/E ,
\label{potential}\end{equation}
with
\begin{equation}
\lambda_\ell  ={g^2\over 32\pi^2} {(eB)^2\over m^4_W}\left[1 -{4\over 3}\ln\left({eB\over 6m^2_W} +\delta_\ell \right)\right]\chi,
\label{lambda}\end{equation}
where $\ell = e,\mu,\tau$. 

\section{Resonance condition and oscillation length}
\label{5}

Our main results, Eqs. (\ref{mu2}) - (\ref{lambda}), have significant consequences on neutrino oscillations of the type 
$\nu_e\leftrightarrow \nu_{\mu,\tau}$. While the original MSW effect \cite{Wolfenstein:1977ue,Wolfenstein:1979ni,Mikheev:1986gs} showed that the presence of a medium alone can produce neutrino oscillations, in this work we will explore the role of a lone magnetic field in neutrino oscillations. 
The mixing angle $\theta_B$ in a magnetic field is determined by the following
\begin{equation}
\sin^2 2\theta_B={\sin^2 2\theta_{1i}\over{\left[\cos2\theta_{1i}-{2E(V_e-V_\ell)\over \Delta m^2_{i1}}\right]^2}+\sin^2 2\theta_{1i}}
\label{mixingangle}
\end{equation}
where $\Delta m^2_{i1}$ and $\theta_{1i}$ are the squared-mass splitting and vacuum mixing angle in the $\nu_e$, $\nu_\ell$ system, with $\ell=\mu, \tau$ and $i=2,3$ for $\ell = \mu, \tau$, and
$V_e$ and $V_\ell$ are the $\nu_e$ and $\nu_\ell$ effective potentials in a magnetic field obtained in Eq. (\ref{potential}). Resonant oscillations will occur if
\begin{equation}
{\Delta m^2_{i1} }\cos 2\theta_{i1} = 2E(V_e-V_\ell)
\label{resonance}
\end{equation}
and, using Eq. (\ref{potential}), we find that the resonance conditions is
\begin{equation}
{\Delta m^2_{i1} }\cos 2\theta_{i1} \simeq 2(\lambda_e-\lambda_\ell) p^2_\perp,
\label{resonance2}
\end{equation}
where we neglected the smaller term proportional to $\mu$ in (\ref{potential}) and
\begin{equation}
\lambda_e-\lambda_\ell= {g^2\over 24\pi^2} {(eB)^2\over m^4_W}\chi\ln\left({eB+6m^2_\ell\over eB+6m^2_e}\right).
\label{lambdas}
\end{equation}
We take $\chi \sim 1$, the newest and most accurate values of ${\Delta m^2_{i1} }$ and $\cos 2\theta_{i1}$ \cite{Tanabashi:2018oca}, and in Figures 1 and 2 plot the value of $ p_\perp$ for which resonance occurs as a function of $B$, for 
$0<B\le4B_e$, where $B_e=m^2_e/e=4.4\times 10^{13} \,{\rm G}$. Figure 1 shows our result for $\nu_e$-$\nu_\mu$ oscillations, Figure 2  shows it for $\nu_e$-$\nu_\tau$ oscillations. 
\begin{figure}[b]
\centerline{\includegraphics[width=10.0cm]{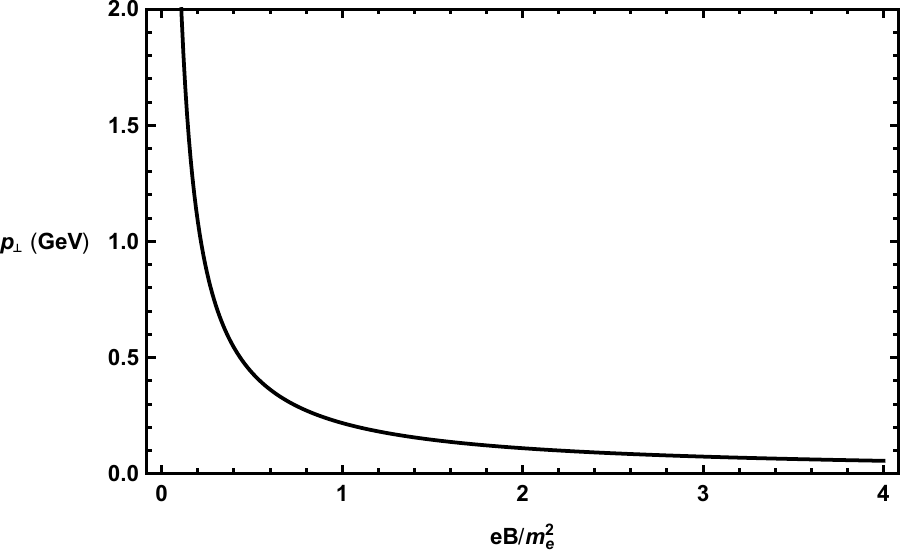}}
\caption{
Plot of $p_\perp$ for which resonance occurs as a function of $B$, in the case of $\nu_e$-$\nu_\mu$ oscillations.  $p_\perp$ is in GeV and  
$0<B\le4B_e$.
\label{fig1}
           }
\end{figure}
\begin{figure}[b]
\centerline{\includegraphics[width=10.0cm]{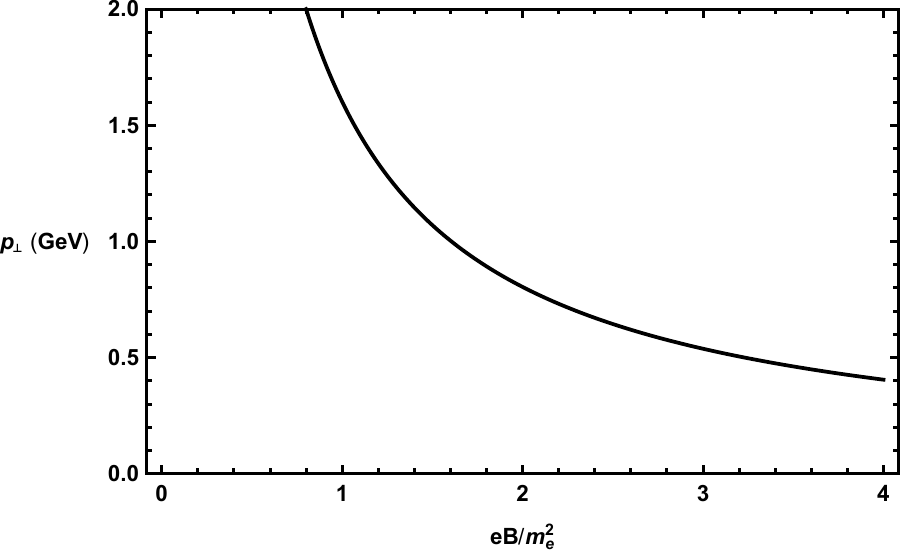}}
\caption{
Plot of $p_\perp$ for which resonance occurs as a function of $B$, in the case of $\nu_e$-$\nu_\tau$ oscillations.  $p_\perp$ is in GeV and  
$0<B\le4B_e$.
\label{fig2}
           }
\end{figure}

Neutrino oscillatory behavior is prominent when $L\sim L^{osc}$, where $L$ is the distance travelled inside the magnetic field and $L^{osc}$ is the oscillation length . The oscillation length for neutrinos in a
magnetic field is 
\begin{equation}
L^{osc}={L^{osc}_0\over\sqrt{{\left[\cos2\theta_{1i}-{2(\lambda_e-\lambda_\ell) p^2_\perp)\over \Delta m^2_{i1}}\right]^2}+\sin^2 2\theta_{1i}}}
\label{losc}
\end{equation}
where $L^{osc}_0$ is the vacuum oscillation length, given by
\begin{equation}
L^{osc}_0={4\pi E\over \Delta m^2_{i1}}=2.5 {E\over  \Delta m^2_{i1}}\,\, {\rm meter}
\label{losc0}
\end{equation}
where the neutrino energy $E$ is in MeV and $\Delta m^2_{i1}$ is in eV$^2$. The neutrino energy is easily related to  $p_\perp$ by assuming that the three components of the neutrino momentum are of similar magnitude, $p_1\simeq p_2\simeq p_3$, which leads to $E\sim \sqrt{3\over 2}p_\perp$. The oscillation length at resonance is
\begin{equation}
L^{osc}_R={L^{osc}_0\over \sin2\theta_{1i}},
\label{loscr}
\end{equation}
and, in the case of $\nu_e$-$\nu_\mu$ oscillations when $B=B_e$ and $E\sim 270\,{\rm MeV}$, we find $L^{osc}_R\sim10^4\, {\rm Km}$.
For  $\nu_e$-$\nu_\mu$ oscillations, $B=4B_e$, and $E\sim 70\,{\rm MeV}$, the oscillation length is  $L^{osc}_R\sim2.5\times10^3\, {\rm Km}$. Many astrophysical sites such as 
supernovae, neutron stars, white dwarfs and magnetars have magnetic fields as large as $B_e$ or larger, even larger by more than one order of magnitude, and therefore $\nu_e$-$\nu_\mu$ resonant oscillations in magnetic field are likely to occur in the proximity of such objects.

\end{document}